\documentclass[12pt]{iopart}

\usepackage{iopams}
\usepackage{braket}
\usepackage{amsfonts}
\usepackage{color}
\usepackage[dvipsnames]{xcolor}
\usepackage{amssymb}
\usepackage{mathrsfs}
\usepackage{graphicx}
\usepackage{lmodern}
\usepackage{lineno}
\usepackage{hyperref}
\usepackage[normalem]{ulem}

\newcommand{\beq}{\begin{eqnarray}}
\newcommand{\eeq}{\end{eqnarray}}

\newcommand{\nn}{\nonumber}

\def\keywords#1{\vspace{10pt}
     \begin{indented}
     \item[]\rm Keywords: #1\par
     \end{indented}}

\begin{document}



\title{Deformation quantization of constrained systems: a group averaging approach}
\author{Jasel Berra--Montiel$^{1,2}$ and 
Alberto Molgado$^{1,2}$}

\address{$^{1}$ Facultad de Ciencias, Universidad Aut\'onoma de San Luis 
Potos\'{\i} \\
Campus Pedregal, Av. Parque Chapultepec 1610, Col. Privadas del Pedregal, San
Luis Potos\'{\i}, SLP, 78217, Mexico}
\address{$^2$ Dual CP Institute of High Energy Physics, Mexico}

\eads{\mailto{\textcolor{blue}{jasel.berra@uaslp.mx}},\ 
\mailto{\textcolor{blue}{alberto.molgado@uaslp.mx}}\ 
}


\begin{abstract}

Motivated by certain concepts introduced by 
the Refined Algebraic Quantization formalism
for constrained systems which has been successfully  applied within the context of 
Loop Quantum Gravity, in this paper we propose
a phase space implementation of the Dirac quantization formalism to appropriately include
systems with constraints.  In particular, 
we propose a physically prescribed Wigner 
distribution which allows the definition of a
well-defined inner product by judiciously introducing a star version of the group averaging of the constraints.  This star group
averaging procedure is obtained by 
considering the star-exponential of the 
constraints and then integrating with respect to 
a suitable measure.  In this manner,  
the proposed physical Wigner distribution
explicitly 
solves the constraints by including 
generalized functions on phase 
space.  Finally, in order to illustrate our approach, our proposal is tested in a couple of simple examples which recover standard results found 
in the literature.
\end{abstract}

\keywords{Deformation quantization, star-product, constrained systems, group averaging}
\ams{81S30, 53D55, 70H45}


\section{Introduction}

Constrained classical systems, sometimes referred to as singular systems, appear whenever the dynamics associated to them cannot be described by choosing an arbitrary set of initial values. In consequence, the appearance of constraints implies that the dynamical formulation of a system 
involve some redundancy. This means that the number of distinguishable physical solutions is thus smaller than the number of initial data required by the set of differential equations which characterize the evolution of a system. In addition, the presence of certain type of constraints, called first class constraints, play a special role since they not only restrict initial values but also are associated to the generators of gauge transformations. A gauge transformation is a map that exhibits an underlying symmetry in the space of solutions of the equations of motion by identifying different mathematical solutions as distinct representations of the same physical result \cite{Henneaux,Rothe}. Notoriously, our current comprehension of all of the theories that describe the fundamental interactions, including the gravitational case, correspond to gauge theories. This last reason makes the study of systems possessing gauge symmetries so relevant in physics.

As it is well known, the quantization of such systems can be traced back  to the roots of the classical analysis of singular Lagrangians, where the basic outline was introduced by Dirac \cite{Dirac}. This formalism, known as the Dirac quantization program for constrained systems, has remained at the forefront of the quantum gravity research as the nature of gauge transformations associated to the gravitational field render gauge fixing techniques notably difficult to apply and, besides, the 
non-renormalizability of the theory has prevented to employ perturbative approaches such as the covariant path integral quantization \cite{Witt,Tate}. Despite the wide applicability of Dirac's program to many situations, one faces certain difficulties while quantizing constrained systems. First, the introduction of a Hilbert space with a suitable measure where operators of physical interest are to be defined poses a non-trivial challenge. Then, even after one has established a Hilbert space, it remains the difficult task of solving the constraints to distinguish the physical states and for which an appropriate inner product is well-defined. 
Both issues result very problematic in practice since, generically, even for systems with a finite number of degrees of freedom, the solutions of the constraints do not lie on the proposed initial Hilbert space. In order to resolve these subtleties, some refinements of the Dirac quantization
program have been introduced through the years, including geometric quantization \cite{Woodhouse,Tuynman}, BRST methods \cite{Henneaux,Becchi}, integral coherent quantization methods \cite{Gazeau,GazeauC}, the projector-operator approach~\cite{klauder1,klauder2}, and algebraic quantization 
\cite{Tate,TateE}, among others.  

In this paper, we propose to analyze the quantization of constrained systems within the deformation quantization formalism. Deformation quantization, also referred to as phase space quantum mechanics by many authors, consists in a general method to pass from a classical system, encoded by the Poisson structure, to its corresponding quantum system \cite{Bayen,Flato}. The main feature of this quantization approach lies in the emphasis acquired by the algebra of quantum observables, which is not given by a family of operators on a Hilbert space. Instead, these observables correspond to smooth real or complex valued functions defined on the phase space, where the usual point-wise commutative product is replaced by a non-commutative product, denoted as the star-product. Consequently, this star-product induces a deformation of the Poisson bracket in such a manner that it contains all the information related to the commutators between self-adjoint operators. One crucial element within this formulation resides on the definition of the Wigner distribution, which corresponds to a phase space representation of the density matrix that is responsible for all auto-correlation properties and transition amplitudes of a given quantum mechanical system. Despite the deformation quantization formalism have undoubtedly provided important contributions not only 
in pure mathematics \cite{Kontsevich,Reichert}, but it has 
also prove to be a reliable technique in the understanding of many physical quantum systems \cite{Waldmann,Curtright}, including recently certain aspects of the loop representation of quantum cosmology and quantum gravity \cite{DQPoly,DQScalar}.  
However, it is important to mention that the application of the 
deformation quantization formalism to constrained systems has been poorly developed \cite{Antonsen,Batalin,Hori}, (see also 
references~\cite{Compean,compean2} for applications in constrained models associated to Field theory and Cosmology, respectively). 

Following some ideas formulated in \cite{RAQ,Giulini,Giulini2}, the aim of this paper is to construct a physical Wigner distribution and the physical inner product by judiciously 
implementing  
a star group averaging of the constraints which, as we will demonstrate below, consists in integrating the star-exponential of the constraints with respect to some appropriate measure. This integral defines a map which takes the kinematic (unphysical) Wigner distribution to the physical Wigner distribution. Then, since the physical Wigner distribution corresponds to the phase space representation of the density operator, the physical inner product is obtained by integrating this physical Wigner distribution over the whole phase space. 

The paper is organized as follows, in section \ref{sec:DQ}, we 
briefly introduce the basic ideas of deformation quantization. In section \ref{sec:RAQ}, the Refined Algebraic Quantization (RAQ) and the group averaging procedure for systems with constraints are reviewed. In section \ref{sec:DQC}, we obtain the physical Wigner distribution and the physical inner product by implementing the star group averaging technique and we present a couple of examples in order to illustrate our approach. Finally, we introduce some concluding remarks in section \ref{sec:conclu}.         

\section{Deformation quantization }
\label{sec:DQ}
This section is devoted to briefly review the basic ideas of deformation quantization. First, we introduce the Wigner-Weyl quantization procedure on a given set of classical observables. Then, we take advantage of the deformation quantization approach by obtaining a non-commutative 
star-product on the quantum algebra of observables. We encourage the reader to see the more detailed reviews \cite{Bordemann,Blaszak,Gutt,Takhtajan} in order to get a full perspective on the deformation quantization program.

\subsection{The Wigner-Weyl quantization}

The canonical quantization of a classical system on the phase space $\Gamma=\mathbb{R}^{2n}$ can be obtained by establishing a correspondence between the set of classical observables $\mathcal{A}=C^{\infty}(\mathbb{R}^{2n})$, and the set of self-adjoint operators $\mathfrak{U}$ on a Hilbert space $\mathcal{H}$. This correspondence, also known as Bohr's principle, can be achieved by defining a one-to-one mapping $Q_{\hbar}:\mathcal{A}\to\mathfrak{U}$, which depends on a parameter $\hbar>0$ and satisfies the following properties
\begin{equation}
\lim_{\hbar\to 0}\frac{1}{2}Q_{\hbar}^{-1}\left(Q_{\hbar}(f)Q_{\hbar}(g)+Q_{\hbar}(g)Q_{\hbar}(f)\right)=fg, 
\end{equation}
and
\begin{equation}
\lim_{\hbar\to 0}Q_{\hbar}^{-1}\left(\frac{i}{\hbar}\left[ Q_{\hbar}(f),Q_{\hbar}(g)\right]\right) =\left\lbrace f,g \right\rbrace. 
\end{equation}
The association between the classical observables, $f\in C^{\infty}(\mathbb{R}^{2n})$, and their quantum counterparts, $Q_{\hbar}(f)\in\mathfrak{U}$, suggests that the quantization procedure is merely a deformation of the algebraic structures of classical observables realized as an isomorphism of Lie algebras only in the limit $\hbar\to 0$, that is, when the quantum description of physical phenomena turns into classical mechanics.
In the case of a classical system with $n$ degrees of freedom, described within the phase space $\mathbb{R}^{2n}$, with local coordinates $\mathbf{p}=(p_{1},\cdots,p_{n})$ and $\mathbf{q}=(q^{1},\cdots,q^{n})$, the quantization mapping $Q_{\hbar}:\mathcal{A}\to\mathfrak{U}$ means the passage from the Poisson brackets
\begin{equation}
\left\lbrace p_{i},p_{j} \right\rbrace=0, \;\;\left\lbrace q^{i},q^{j} \right\rbrace=0, \;\; \left\lbrace q^{i},p_{j} \right\rbrace=\delta^{i}_{j},   
\end{equation}  
to the commutator of operators
\begin{equation}\label{Hcommutators}
\left[\hat{P}_{i},\hat{P}_{j} \right]\psi=0, \;\; \left[\hat{Q}^{i},\hat{Q}^ {j}\right]\psi=0, \;\; \left[\hat{Q}^{i},\hat{P}_{j} \right]\psi=i\hbar\delta^{i}_{j}\psi, \;\;\; \textrm{for}\;\psi\in D, 
\end{equation}
where $ D\subset\mathcal{H}$, is a dense linear subset of the Hilbert space $\mathcal{H}$, such that the operators $\hat{P}_{i}$, $\hat{Q}^{j}$ and their commutators yield self-adjoint operators. These relations are known as the Heisenberg commutation relations for $n$ degrees of freedom and the algebraic structure underlying them is given by the so-called Heisenberg algebra. The association $Q_{\hbar}(p_{i})=\hat{P}_i$ and $Q_{\hbar}(q^{j})=\hat{Q}^{j}$, satisfying relations (\ref{Hcommutators}) provides the standard rule for the quantization of most classical systems, and its validity remains widely confirmed by numerous experiments so far.

In order to find an specific expression for the quantization mapping $Q_{\hbar}$, we consider the algebra generated by the strongly continuous unitary operators 
\beq
\hat{U}(\mathbf{u}) 
& = & \exp{\left( \sum_{i=1}^{n}-iu^{i}\hat{P}_{i}/\hbar\right) }=:e^{-i\mathbf{u\hat{P}}/\hbar} \,, 
\nn\\
\hat{V}(\mathbf{v})
& = & 
\exp{\left( \sum_{j=1}^{n}-iv_{j}\hat{Q}^{j}/\hbar\right) }=:e^{-i\mathbf{v\hat{Q}}/\hbar} \,,
\label{UandV}
\eeq
where $\mathbf{u}=(u^{1},\ldots,u^{n})$, and $\mathbf{v}=(v_{1},\ldots,v_{n})$ denote a set of parameters in $\mathbb{R}^{n}$. Then, it follows from (\ref{Hcommutators}) that the unitary operators $\hat{U}(\mathbf{u})$ and $\hat{V}(\mathbf{v})$ satisfy the Weyl commutations relations
\begin{equation}\label{UV}
\hat{U}(\mathbf{u})\hat{V}(\mathbf{v})=e^{i\mathbf{u}\mathbf{v}/\hbar}\hat{V}(\mathbf{v})\hat{U}(\mathbf{u}),
\end{equation}
where $\mathbf{u}\mathbf{v}=\sum_{i=1}^{n}u^iv_{i}$ indicates the usual inner product on $\mathbb{R}^{n}$. The starting point of deformation quantization can be traced back to Weyl's quantization procedure \cite{Weyl}, where it is proposed that given a classical observable on the phase space $f\in C^{\infty}(\mathbb{R}^{2n})$, we can associate to $f$ an operator or quantum observable in the Hilbert space $\mathcal{H}=L^{2}(\mathbb{R}^{n})$ by introducing the following linear map
\begin{equation}
\Phi=C^{\infty}({\mathbb{R}^{2n}})\to\mathcal{L}(\mathcal{H}),
\end{equation}
called the Weyl quantization and it is explicitly stated by the integral
\begin{equation}\label{Weylquant}
\Phi(f)=\frac{1}{(2\pi\hbar)^{n}}\int_{\mathbf{R}^{2n}}\tilde{f}(\mathbf{u},\mathbf{v})\hat{S}(\mathbf{u},\mathbf{v})\,d^{n}\mathbf{u}\,d^{n}\mathbf{v},
\end{equation}
where $\tilde{f}(\mathbf{u},\mathbf{v})$ stands for the inverse Fourier transform
\begin{equation}\label{inverseFourier}
\tilde{f}(\mathbf{u},\mathbf{v})=\mathfrak{F}^{-1}(f)(\mathbf{u},\mathbf{v})=\frac{1}{(2\pi\hbar)^{n}}\int_{\mathbb{R}^{2n}}f(\mathbf{p},\mathbf{q})e^{\frac{i}{\hbar}(\mathbf{up}+\mathbf{vq})}\,d^{n}\mathbf{p}\,d^{n}\mathbf{q},
\end{equation}
and $\hat{S}(\mathbf{u},\mathbf{v})=e^{-\frac{i}{2\hbar}\mathbf{uv}}\hat{U}(\mathbf{u})\hat{V}(\mathbf{v})$. By using (\ref{UandV}) and the inverse Fourier transform (\ref{inverseFourier}), the Weyl quantization map reads
\begin{equation}\label{Weylq}
\Phi(f)\psi(\mathbf{q})=\frac{1}{(2\pi\hbar)^n}\int_{\mathbb{R}^{2n}}f\left(\mathbf{p},\frac{\mathbf{q}+\mathbf{q}'}{2}\right)e^{\frac{i}{\hbar}\mathbf{p}(\mathbf{q}-\mathbf{q}')}\psi(\mathbf{q}')\,d^{n}\mathbf{p}\,d^{n}\mathbf{q}', 
\end{equation}
for $\psi\in L^{2}(\mathbb{R}^{n})$. This means, that the operator $\Phi(f)$ represents an integral operator acting on $\mathcal{H}$, 
\begin{equation}
\Phi(f)\psi(\mathbf{q})=\int_{\mathbb{R}^{n}}K(\mathbf{q},\mathbf{q}')\psi(\mathbf{q}')\,d^{n}\mathbf{q}',
\end{equation}
where the kernel $K(\mathbf{q},\mathbf{q}')$ is given by
\begin{equation}
K(\mathbf{q},\mathbf{q}')=\frac{1}{(2\pi\hbar)^n}\int_{\mathbb{R}^{n}}f\left(\mathbf{p},\frac{\mathbf{q}+\mathbf{q}'}{2}\right)e^{\frac{i}{\hbar}\mathbf{p}(\mathbf{q}-\mathbf{q}')}\,d^{n}\mathbf{p}.
\end{equation}
Properly speaking, this integral should be understood as a limit of Riemann sums with respect to the norm topology on $\mathcal{L}(\mathcal{H})$. Further, in order to extend the Weyl transform to a more general class of functions, such as distributions, it is convenient to allow functions on the Schwartz space $\mathcal{S}(\mathbb{R}^{2n})$ as classical observables, that is, $C^{\infty}$ functions defined on $\mathbb{R}^{2n}$ whose derivatives are rapidly decreasing. Bearing this in mind, and since the Fourier transform maps the Schwartz space onto itself, this means that the Weyl quantizer turns out to be a Hilbert-Schmidt operator acting on $L^{2}({\mathbb{R}^{n}})$, that is, a non-negative, self-adjoint operator with a well-defined (but possible infinite) trace \cite{Hall}. The inverse map 
$\Phi^{-1}:L^2(\mathbb{R}^n)\to
\mathcal{S}(\mathbb{R}^{2n})$, also known as Weyl's inversion formula, can be obtained as
\begin{equation}\label{Winverse}
f(\mathbf{p},\mathbf{q})=\mathrm{tr} \left(\Phi(f)\hat{S}(\mathbf{p},\mathbf{q})^{-1}\right), 
\end{equation}  
for $f\in\mathcal{S}(\mathbb{R}^{2n})$.
Here the trace is taken with respect to any orthonormal basis for $L^{2}(\mathbb{R}^{n})$. Following the terminology of harmonic analysis \cite{Reed}, we will say that 
Weyl's inversion formula (\ref{Winverse}) defines a symbol $f$ from its quantization $\hat{f}$, that is, a phase space representation of the operator $\hat{f}$ acting on $\mathcal{H}$. 

With the Weyl quantization map and its inverse at hand, we are now ready to define the Wigner distribution corresponding to the Hilbert space $\mathcal{H}$. Let $\hat{\rho}$ be a density operator associated to a quantum state $\psi\in\mathcal{H}$, that is, a self-adjoint, positive semi-definite operator with trace one, written as
\begin{equation}\label{density}
\hat{\rho}\varphi(\mathbf{q})=\psi(\mathbf{q})\int_{\mathbb{R}^{n}}\overline{\psi(\mathbf{q}')}\varphi(\mathbf{q'})\,d^{n}\mathbf{q}',
\end{equation}
(or $\ket{\psi}\bra{\psi}$ in Dirac notation), where $\psi$, $\varphi\in\mathcal{H}$. Since the operator $\hat{\rho}$ amounts to a integral operator (\ref{density}), by Weyl's inversion formula (\ref{Winverse}), its corresponding symbol is given by
\begin{equation}\label{Wigner}
\rho(\psi)(\mathbf{p},\mathbf{q})=\int_{\mathbb{R}^{n}}\psi\left(\mathbf{q}+\frac{\mathbf{z}}{2}\right) \overline{\psi \left(\mathbf{q}-\frac{\mathbf{z}}{2} \right)}e^{-\frac{i}{\hbar}\mathbf{zp}}\,d^{n}\mathbf{z}.
\end{equation}   
This is the Wigner distribution associated to the Hilbert space $\mathcal{H}$ and, as one may easily check, it is normalized 
\begin{equation}
\frac{1}{(2\pi\hbar)^{n}}\int_{\mathbb{R}^{2n}}\rho(\psi)(\mathbf{p},\mathbf{q})\,d^{n}\mathbf{p}\,d^{n}\mathbf{q}=1.
\end{equation}
A distinguished property of the Wigner distribution for a quantum state lies on the possibility to acquire negative values on certain regions of phase space and, therefore, it can not be interpreted as a probability density in the standard sense of statistical mechanics, thus it is usually referred to as a quasi-probability distribution in the literature. However, this apparently odd feature impress Wigner distribution 
with a very  relevant feature since it allows to visualize 
the fact that quantum trajectories in phase space that allow negative probability values may serve to characterize joint-correlation functions and entanglement properties within the quantum system~\cite{Curtright,weinbub}.  Moreover, the Wigner distribution can be used to obtain the expectation value of an arbitrary operator by integrating its symbol over the whole phase space,
\begin{equation}\label{expect}
\braket{\psi,\hat{A}\psi}=\frac{1}{(2\pi\hbar)^{n}}\int_{\mathbb{R}^{2n}}\rho(\psi)(\mathbf{p},\mathbf{q})A(\mathbf{p},\mathbf{q})\,d^{n}\mathbf{p}\,d^{n}\mathbf{q}  \,,
\end{equation} 
where the operator $\hat{A}=\Phi(A)$ corresponds to the Weyl transform of the classical function $A(\mathbf{p},\mathbf{q})$. All these properties imply that the Wigner distribution is the closest object we have to a probability
distribution for a quantum system and, since it corresponds to the phase space representation of a density operator, one may argue that the information encompassed by the Wigner distribution is completely equivalent to knowing the wave functions at the standard quantum mechanical 
level~\cite{Curtright}. 

\subsection{The star-product}
The Weyl-Wigner quantization $\Phi:\mathcal{S}(\mathbb{R}^{2n})\to\mathcal{L}(\mathcal{H})$, briefly described in the previous subsection, defines a bilinear map denoted by $\star:\mathcal{S}(\mathbb{R}^{2n})\times\mathcal{S}(\mathbb{R}^{2n})\to\mathcal{S}(\mathbb{R}^{2n})$, and explicitly 
given by
\begin{equation}\label{dstar}
f\star g=\Phi^{-1}\left( \Phi(f)\Phi(g)\right)\,.
\end{equation}
This map is generally called the star-product or the Moyal product. By substituting explicitly the Weyl quantization map in terms of the integral operator (\ref{Weylq}) and using  Weyl's inversion formula (\ref{Winverse}), after some manipulations of Taylor series expansions (see~\cite{Compean},~\cite{Hirshfeld} for further details), the star-product can be written as
\begin{equation}\label{star}
(f\star g)(\mathbf{p},\mathbf{q})=f(\mathbf{p},\mathbf{q})\exp\left[ \frac{i\hbar}{2}\left(\frac{\overleftarrow{\partial}}{\partial\mathbf{q}}\frac{\overrightarrow\partial}{\partial\mathbf{p}}-\frac{\overleftarrow{\partial}}{\partial\mathbf{p}}\frac{\overrightarrow\partial}{\partial\mathbf{q}}\right) \right]g(\mathbf{p},\mathbf{q}),
\end{equation}
where the differential operators $\overleftarrow{\partial}/\partial\mathbf{q}$, $\overleftarrow{\partial}/\partial\mathbf{p}$ act on the left, and the differential operators $\overrightarrow{\partial}/\partial\mathbf{q}$, $\overrightarrow{\partial}/\partial\mathbf{p}$ act on the right. This product defines an associative formal deformation of the ordinary point-wise product of functions on $\mathbb{R}^{2n}$, since we have
\begin{equation}
\left( f\star(g\star h)\right)(\mathbf{p},\mathbf{q})=\left( (f\star g)\star h \right)(\mathbf{p},\mathbf{q}),  
\end{equation}
and
\begin{equation}
(f\star g)(\mathbf{p},\mathbf{q})=(fg)(\mathbf{p},\mathbf{q})+\frac{i\hbar}{2}\left\lbrace f,g \right\rbrace (\mathbf{p},\mathbf{q})+O(\hbar^{2})\;\;\textrm{as}\;\hbar\to 0,
\end{equation}
where $\left\lbrace f,g \right\rbrace$ stands for the canonical Poisson bracket on the phase space $\mathbb{R}^{2n}$.

Analogously to the Hilbert space formulation, within the deformation quantization approach the spectral properties of operators are encoded by star-genvalues equations. In particular, a stationary Wigner distribution (\ref{Wigner}) satisfies
\begin{equation}\label{star-energy}
\mkern-55mu H(\mathbf{p},\mathbf{q})\star\rho(\psi)(\mathbf{p},\mathbf{q})=H\left( \mathbf{p}-\frac{i\hbar}{2}\frac{\overrightarrow{\partial}}{\partial\mathbf{q}},\mathbf{q}+\frac{i\hbar}{2}\frac{\overrightarrow{\partial}}{\partial\mathbf{p}}\right)\rho(\psi)(\mathbf{p},\mathbf{q})=E\rho(\psi)(\mathbf{p},\mathbf{q}), 
\end{equation} 
where $H(\mathbf{p},\mathbf{q})$ is the
classical Hamiltonian function and $E$ corresponds to the energy eigenvalues of $\hat{H}\psi=E\psi$ in the Schr\"odinger representation \cite{Curtright}. Moreover, excluding the degenerate case, that is any multiplicity on the star-genvalues, it follows
\begin{equation}
\rho(\psi)\star H(\mathbf{p},\mathbf{q})\star\rho(\psi)=E\rho(\psi)\star \rho(\psi)=H(\mathbf{p},\mathbf{q})\star\rho(\psi)\star \rho(\psi),
\end{equation}
which implies that $\rho(\psi)\star\rho(\psi)=\frac{1}{\hbar^{n}}\rho(\psi)$. These projective features of the star-genfunctions amount to a complete characterization of the Wigner distribution \cite{Zachos}. Finally, the star-product satisfies the following cyclic phase space integration property
\begin{equation}\label{cyclic}
\int_{\mathbb{R}^{2n}}\rho(\psi)(\mathbf{p},\mathbf{q})\star A(\mathbf{p},\mathbf{q})\,d^{n}\mathbf{p}\,d^{n}\mathbf{q}= \int_{\mathbb{R}^{2n}}\rho(\psi)(\mathbf{p},\mathbf{q})A(\mathbf{p},\mathbf{q})\,d^{n}\mathbf{p}\,d^{n}\mathbf{q}   \,,
\end{equation} 
for any classical function $A(\mathbf{p},\mathbf{q})$ defined on phase space. This means that the star-genvalue equations (\ref{star-energy}) are necessary, but not sufficient to specify the Wigner function for a quantum system \cite{Curtright,Hirshfeld}.

\section{Refined algebraic quantization and group averaging}
\label{sec:RAQ}

Refined Algebraic Quantization (RAQ) \cite{RAQ,Giulini,Giulini2}, also known as the Rieffel induction procedure \cite{Rieffel}, consists of a well--defined mathematical framework developed in order to quantize constrained classical systems. The method itself has already been successfully applied to several examples like linearized gravity on symmetric backgrounds \cite{Higuchi}, various minisuperspace models \cite{Marolf,Marolf1}, diffeomorphism invariant theories such as General Relativity \cite{Thiemann}, and certain finite dimensional models associated to General Relativity~\cite{gomberoff,Louko0,Louko1,Louko2,Louko3}. In this section, we review the RAQ scheme for systems with gauge symmetries.

Consider a classical system with first class constraints $(C_{I})_{I\in\mathcal{I}}=0$,  defined in a phase space given by a real symplectic manifold, where the label set $\mathcal{I}$ denotes the number of constraints (which in principle can be finite or infinite). The essential idea of RAQ is to turn the classical constraints $C_{I}$  into linear operators $\hat{C}_{I}$, defined on an auxiliary Hilbert space usually referred as $\mathcal{H}_{kin}$, and then look for states $\psi\in\mathcal{H}_{kin}$ that are annihilated by the constraints, that is, $\hat{C}_{I}\psi=0$, in terms of generalized eigenvectors in the algebraic dual of some dense subspace $\mathcal{D}\subset\mathcal{H}_{kin}$. The kinematic Hilbert space $\mathcal{H}_{kin}$, is supposed to implement the adjointness and canonical commutation relations of the kinematic degrees of freedom. These  degrees of freedom do not correspond to physical states, since generically the point zero does not lie in the discrete part of the spectrum of $\hat{C}_{I}$. In some sense, the task of $\mathcal{H}_{kin}$ is to provide a common dense domain $\mathcal{D}\subset\mathcal{H}_{kin}$ to all operators $\hat{C}_{I}$, such that their adjoints remain densely defined, that is, the quantum constraints $\hat{C}_{I}$ represent a set of closable operators.

Let us now consider the set of kinematic observables $\mathcal{O}_{kin}$, that is, the set of all self-adjoint operators on $\mathcal{H}_{kin}$ with $\mathcal{D}_{kin}$ as a common dense domain. From $\mathcal{O}_{kin}$ we can define the algebra of physical observables as
\begin{equation}
\mathcal{O}_{phys}=\left\lbrace \hat{O}\in \mathcal{O}_{kin}: [\hat{C}_{I},\hat{O}]=0 \right\rbrace.
\end{equation}
Typically, the space $\mathcal{D}_{kin}$ is equipped with a nuclear topology, namely a topology generated by a countable family of seminorms, which in principle is different to the topology inherited from the Hilbert space $\mathcal{H}_{kin}$. Then, it follows that the topological dual, that is, the space of continuous linear functionals $\mathcal{D}'_{kin}$ contains $\mathcal{H}_{kin}$, thus we have the next topological inclusion
\begin{equation}
\mathcal{D}_{kin}\subset\mathcal{H}_{kin}\subset\mathcal{D}'_{kin},
\end{equation} 
also called a Gelfand triple \cite{Gelfand}. Since the definition of the Gelfand triple requires the use of the nuclear topology coming from $\mathcal{D}_{kin}$, which does not seem to be based on any physical guiding principle \cite{RAQ}, we instead consider the algebraic dual $\mathcal{D}^{*}_{kin}$, that is, the space of all linear functionals on $\mathcal{D}_{kin}$ equipped with the relative topology induced from $\mathcal{H}_{kin}$. The topology defined on $\mathcal{D}^{*}_{kin}$ is naturally given by the weak $^*$-topology of point-wise convergence. Then, as we can regard $\mathcal{H}_{kin}$ as a subset of $\mathcal{D}^{*}_{kin}$, we immediately obtain the next inclusion 
\begin{equation}
\mathcal{D}_{kin}\subset\mathcal{H}_{kin}\subset\mathcal{D}^{*}_{kin},
\end{equation}
which, as an abuse of language, is also called a Gelfand triple in the literature.

The main reason to consider the previous structures is to look for solutions to the constraints $\hat{C}_{I}\psi=0$, for $\psi\in\mathcal{H}_{kin}$ and for all $I\in\mathcal{I}$. Since, as mentioned before, usually the point zero is not part of the discrete spectrum of the constraints, one must allow for generalized functions in the algebraic dual $f\in\mathcal{D}^{*}_{kin}$, which satisfy
\begin{equation}
(\hat{C}_{I}f)(\phi):=f(\hat{C}_{I}\phi)=0,\;\;\textrm{for all }\phi\in\mathcal{D}_{kin}, 
\end{equation}   
let us call $\mathcal{D}^{*}_{phys}\subset\mathcal{D}^{*}_{kin}$ the space of solutions of the constraints. Now, the idea is to select a subspace $\mathcal{H}_{phys}$ of $\mathcal{D}^{*}_{phys}$ and equip it with a Hilbert space topology. The reason to choose a subset $\mathcal{H}_{phys}$ instead of all $\mathcal{D}^{*}_{phys}$, is the appearance of unbounded operators on the algebra of observables, which would not be defined everywhere on $\mathcal{D}^{*}_{phys}$. In order to study the physical states and the representations of the algebra $\mathcal{O}_{phys}$ on $\mathcal{H}_{phys}$, we will focus on the Gelfand triple
\begin{equation}
\mathcal{D}_{phys}\subset\mathcal{H}_{phys}\subset\mathcal{D}^{*}_{phys}.
\end{equation} 

Our next step is to introduce an inner product on the physical Hilbert space $\mathcal{H}_{phys}$. A systematic procedure is given by the definition of the so-called rigging map \cite{Giulini},
\begin{equation}
\eta:\mathcal{D}_{kin}\to\mathcal{D}^{*}_{phys}
\end{equation}
which satisfy the next properties:
\begin{enumerate}
\item It provides a positive semidefinite sesquilinear form
\begin{equation}
\braket{\eta(\phi),\eta(\phi')}_{phys}:=\eta(\phi')(\phi), \;\;\textrm{for all }\phi,\phi'\in\mathcal{D}_{kin}.
\end{equation}
\item For any observable $\hat{O}\in \mathcal{O}_{phys}$, the rigging map preserves the space of solutions
\begin{equation}
\hat{O}\eta(\phi)=\eta(\hat{O}\phi),\;\;\textrm{for all }\phi\in\mathcal{D}_{kin},
\end{equation}
\end{enumerate}
With these properties at hand, a physical inner product for $\psi=\eta(\phi)$, $\psi'=\eta(\phi')$ can be defined as
\begin{equation}\label{inner}
\mkern-55mu \braket{\psi,\hat{O}\psi'}_{phys}=\braket{\eta(\phi),\hat{O}\eta(\phi')}_{phys}=\eta(\hat{O}\phi')(\phi)=\eta(\phi')(\hat{O}\phi)=\braket{\hat{O}\psi,\psi'}_{phys}.
\end{equation}
Therefore, the existence of a rigging map guarantees by construction a
well-defined inner product. A frequent way to construct such a rigging map $\eta$, is through the so-called group averaging procedure. Since the constraints operators $\hat{C}_{I}$, are self-adjoint and form a Lie algebra $\mathfrak{g}$, for each of the constraints $\hat{C}_{I}\in (\hat{C}_{I})_{I\in\mathcal{I}}$  we consider the rigging map by means of the spectral theorem
\begin{equation}\label{rigging}
\eta(\phi):=\int_{G}\exp(it^{I}\hat{C}_{I})\phi\,d\mu (t),
\end{equation}
where $t^{I}$ denotes a set of parameters, and $G$ stands for the Lie group associated to the Lie algebra $\mathfrak{g}$ (most precisely, the connected component of the group), while $d\mu(t)$ represents the invariant Haar measure on $G$ depending on the set of parameters $t^{I}$. According to the group averaging 
procedure~(\ref{rigging}), the physical inner product (\ref{inner}), is given by
\begin{equation}\label{GA}
\braket{\psi,\psi'}_{phys}=\eta(\phi)(\phi')=\int_{G}\braket{\exp(it^{I}\hat{C}_{I})\phi,\phi'}_{kin}\,d\mu(t),
\end{equation}
for $\psi=\eta(\phi)$, $\psi'=\eta(\phi')$. Note that the expression (\ref{GA}) only makes sense when the integral converges absolutely and, formally, it satisfies $\eta(\phi)(\hat{C}_{I}\phi')=0$. Finally, notice that in the case of an Abelian constraint algebra, from the group averaging formula (\ref{GA}), a reasonable rigging map reads 
\begin{equation}
\eta(\phi)=\prod_{I\in\mathcal{I}}\delta(\hat{C}_{I})\phi,
\end{equation}  
where the delta distributions of the operator constraints are defined using the functional calculus associated to the spectral theorem \cite{Reed,Master}. In case the constraints do not form a Lie algebra, that is, if structure functions appear rather than structure constants, the group averaging procedure is not longer valid since the algebra can not be exponentiated any longer. In this situation, a direct integral decomposition of the constraints such as the Master Constraint Program generalizes RAQ, and thus it turns out to be applicable in cases where the gauge group is not compact or even infinite dimensional  \cite{Master}.

\section{Deformation quantization of constrained systems}
\label{sec:DQC}
By means of the RAQ procedure and the group averaging technique analyzed previously, in this section we determine the physical Wigner distribution associated to a quantum system subject to a set of first class constraints $C_{I}=0$. Then, within the deformation quantization approach, we prove that the physical inner product is completely characterized by the 
star-exponential of the constraints. We conclude this section illustrating the proposed method with a couple of examples.

\subsection{The physical Wigner distribution}

In order to obtain the physical Wigner distribution associated to a generic quantum
constrained system, we take as the kinematic Hilbert space the space of 
square integrable functions, $\mathcal{H}_{kin}=L^{2}(\mathbb{R}^{n})$. As already mentioned in the former section, the purpose of the kinematic Hilbert space is to implement the adjointness and commutation relations among the operators acting on $\mathcal{H}_{kin}$. The question we want to address is then how to formulate the physical inner product, defined through the group averaging procedure (\ref{GA}) in terms of a phase space integration of an appropriate Wigner distribution as depicted in (\ref{expect}). First, one only needs to consider the phase space function corresponding to the non-diagonal density operator $\hat{\rho}(\phi,\phi')$ for the 
kinematic quantum states $\phi,\phi'\in\mathcal{D}_{kin}\subset\mathcal{H}_{kin}$, that is, a self-adjoint and positive semi-definite operator written as
\begin{equation}\label{nondensity}
\hat{\rho}(\phi,\phi')\varphi(\mathbf{q})=\phi(\mathbf{q})\int_{\mathbb{R}^{n}}\overline{\phi'(\mathbf{q})}\varphi(\mathbf{q}')d^{n}\mathbf{q'},
\end{equation} 
(or $\ket{\phi}\bra{\phi'}$ in Dirac notation), where $\varphi\in\mathcal{H}_{kin}$. From (\ref{nondensity}), we can observe that the operator $\hat{\rho}(\phi,\phi')$ is an integral operator, then by Weyl's inversion formula (\ref{Winverse}) its corresponding symbol reads
\begin{equation}\label{dWigner}
\rho(\phi,\phi')(\mathbf{p},\mathbf{q})=\int_{\mathbb{R}^{n}}\phi\left(\mathbf{q}+\frac{\mathbf{z}}{2}\right) \overline{\phi' \left(\mathbf{q}-\frac{\mathbf{z}}{2} \right)}e^{-\frac{i}{\hbar}\mathbf{zp}}\,d^{n}\mathbf{z}.
\end{equation}
This expression will be called the kinematic Wigner distribution, namely a non-diagonal Wigner distribution defined in terms of 
quantum states in $\mathcal{D}_{kin}\subset\mathcal{H}_{kin}$ \cite{diagonal}. One may easily check that a set of wave functions which are orthonormal $\left\lbrace \phi_{m} \right\rbrace $, the kinematic 
Wigner distribution satisfies the relation
\begin{equation}
\frac{1}{(2\pi\hbar)^{n}}\int_{\mathbb{R}^{2n}}\rho(\phi_{m},\phi_{m'})(\mathbf{p},\mathbf{q})\,d^{n}\mathbf{p}\,d^{n}\mathbf{q}=\delta_{mm'}  \,.
\end{equation}
The next step is to find the phase-space function associated to the rigging map $\eta$ constructed through the group averaging proposal (\ref{rigging}). Since the star-product obtained in (\ref{dstar}) defines a homomorphism between the classical observables, $C^{\infty}(\mathbb{R}^{2n})$, and linear operators $\mathcal{L}(\mathcal{H}_{kin})$, this means that the symbol corresponding to the formal operator
\begin{equation}
\exp(it^{I}\hat{C}_{I})=1+it^{I}\hat{C}_{I}+\frac{i^{2}}{2!}(t^{I}\hat{C}_{I})(t^{J}\hat{C}_{J})+\cdots,
\end{equation}
is given by the star-exponential \cite{Bayen1,Bayen2}
\begin{equation}
\exp_{\star}(it^{I}C_{I})(\mathbf{p},\mathbf{q})=1+it^{I}C_{I}(\mathbf{p},\mathbf{q})+\frac{i^{2}}{2!}(t^{I}C_{I})\star(t^{J}C_{J})(\mathbf{p},\mathbf{q})+\cdots,
\end{equation}
for each of the constraints operators $\hat{C}_{I}=\Phi(C_{I})\in (\hat{C}_{I})_{I\in\mathcal{I}}$, related to the classical constraints $C_{I}(\mathbf{p},\mathbf{q})$ by means of the Weyl quantization 
map~(\ref{Weylq}). Using the integral property of the Wigner distribution (\ref{expect}), and the cyclic phase space behavior of the star-product (\ref{cyclic}), the physical inner product obtained in (\ref{GA}) can be expressed as
\begin{equation}\label{Winner}
\braket{\psi,\psi'}_{phys}=\frac{1}{(2\pi\hbar)^{n}}\int_{\mathbb{R}^{2n}\times G}\rho(\phi,\phi')\star\exp_{\star}(it^{I}C_{I})\,d\mu(t)\,d^{n}\mathbf{p}\,d^{n}\mathbf{q}.
\end{equation} 
Bearing this in mind, for an arbitrary operator $\hat{O}=\Phi(O)\in\mathcal{O}_{phys}$ in the algebra of physical observables, the inner product (\ref{inner}) also satisfies
\begin{equation}\label{Wobs}
\mkern-20mu \braket{\psi,\hat{O}\psi'}_{phys}=\frac{1}{(2\pi\hbar)^{n}}\int_{\mathbb{R}^{2n}\times G}\rho(\phi,\phi')\star\exp_{\star}(it^{I}C_{I})\star O(\mathbf{p},\mathbf{q})\,d\mu(t)\,d^{n}\mathbf{p}\,d^{n}\mathbf{q}.
\end{equation}
The preceding expression suggests that the physical non-diagonal Wigner distribution is given by
\begin{equation}\label{Wignerphys}
\rho_{phys}(\psi,\psi')(\mathbf{p},\mathbf{q})=\int_{G}\rho(\phi,\phi')\star\exp_{\star}(it^{I}C_{I})\,d\mu(t),
\end{equation}
where $\rho_{phys}(\psi,\psi')$ would be realized as a generalized function on phase space, and is obtained by star group averaging the kinematic Wigner distribution $\rho(\phi,\phi')$ along the group generated by the exponentiation of the constraint algebra. Of course, one must check case by case whether the Wigner distribution and the formal star-exponential are well defined and, in particular, if the integral formulas converge in an appropriate sense.  Moreover, since the physical Wigner distribution $\rho_{phys}(\psi,\psi')$ corresponds to the phase space representation of a physical density operator, this implies that the physical quantum states $\psi,\psi'\in\mathcal{H}_{phys}$, can be obtained by analyzing the physical inner product defined on (\ref{Winner}) (see examples below).
The relation among observables and the physical inner product proposed in (\ref{Wobs}), shows that $\rho_{phys}(\psi,\psi')$ also satisfies
\begin{equation}\label{Mconstraint}
\braket{\psi,\hat{C}_{I}\psi'}_{phys}=\frac{1}{(2\pi\hbar)^{n}}\int_{\mathbb{R}^{2n}}\rho_{phys}(\phi,\phi')\star C_{I}\,d^{n}\mathbf{p}\,d^{n}\mathbf{q}=0,
\end{equation}
which means that the constraints are solved by the physical Wigner distribution. Expression (\ref{Mconstraint}) completely characterize the gauge invariance of the Wigner distribution and, in principle, it can be used to analyze a quantum constrained system within the deformation quantization perspective without relying at all on the wave function approach.

Notice that in the case of an Abelian set of constraints, the physical Wigner distribution takes the form
\begin{equation}
\mkern-50mu\rho_{phys}(\psi,\psi')(\mathbf{p},\mathbf{q})=\int_{G}\rho(\phi,\phi')\star\exp_{\star}\left( \sum_{I\in\mathcal{I}}it^{I}C_{I}\right) \,d\mu(t)=\rho(\phi,\phi')\star\prod_{I\in\mathcal{I}}\delta_{\star}(C_{I}),
\end{equation}
where we have introduced the Dirac star-delta distribution \cite{Compean} defined as 
\begin{equation}
\delta_{\star}(C_{I}):=\int_{G}\exp_{\star}(it^{I}C_{I})\,d\mu(t).
\end{equation}
A noteworthy feature of the Wigner distribution obtained in (\ref{Wignerphys}) is that, 
whenever the star-exponential of the constraints satisfies the relation $\exp_{\star}(it^{I}C_{I})=\exp(it^{I}C_{I})$, the physical Wigner distribution reads
\begin{equation}
\rho_{phys}(\psi,\psi')(\mathbf{p},\mathbf{q})=\int_{G}\rho(\phi,\phi')\star\exp(it^{I}C_{I})\,d\mu(t)=\rho(\phi,\phi')\delta(C_{I}).
\end{equation}
This last result completely agrees with the Wigner distribution obtained by using integral quantization methods in terms of operator-valued integrals as long as the constraints occur to be real-valued and continuously differentiable functions defined on the configuration space \cite{Gazeau}.

\subsection{Examples}
In this subsection we present a couple of examples in order to illustrate how the physical Wigner distribution can be obtained through the method of star group averaging.

\subsubsection{Single constrained momentum.}

As a first test case, let us consider a non-relativistic particle in two dimensions subject to the classical constraint $p_{1}=0$, this means, that the associated gauge transformations are given by the group of translations in the 
$q_1$-direction, i.e, in this case $G=\mathbb{R}$. Let us consider the kinematic Hilbert space to be $\mathcal{H}_{kin}=L^{2}(\mathbb{R}^{2})$, which allows to implement the adjointness and canonical commutation relation among the basic kinematic observables $q_{1}, q_{2}, p_{1}, p_{2}$. According to (\ref{Wignerphys}), the non-diagonal physical Wigner distribution reads  
\begin{equation}
\rho_{phys}(\psi,\psi')=\int_{\mathbb{R}^{2}\times\mathbb{R}}\phi\left(\mathbf{q}+\frac{\mathbf{z}}{2}\right) \overline{\phi' \left(\mathbf{q}-\frac{\mathbf{z}}{2} \right)}e^{-\frac{i}{\hbar}\mathbf{zp}}\star e_{\star}^{\frac{i}{\hbar}tp_{1}}\,dt\,d^{2}\mathbf{z},
\end{equation}
for $\phi, \phi'\in\mathcal{H}_{kin}$, $t\in\mathbb{R}$ and $dt$ an invariant measure on $G=\mathbb{R}$. Notice that the star-exponential $e_{\star}^{\frac{i}{\hbar}tp_{1}}=e^{\frac{i}{\hbar}tp_{1}}$, this means that the physical Wigner distribution is given by the following generalized function
\begin{equation}
\rho_{phys}(\psi,\psi')=\int_{\mathbb{R}^{2}}\phi\left(\mathbf{q}+\frac{\mathbf{z}}{2}\right) \overline{\phi \left(\mathbf{q}-\frac{\mathbf{z}}{2} \right)}e^{-\frac{i}{\hbar}\mathbf{zp}}\star\delta(p_{1})\,d^{2}\mathbf{z}.
\end{equation}
The physical Wigner distribution can be used to obtain the physical inner product by integrating it over the phase space as
\begin{equation}
\braket{\psi,\psi'}_{phys}=\frac{1}{(2\pi\hbar)^{2}}\int_{\mathbb{R}^{4}}\rho_{phys}(\psi,\psi')(\mathbf{p},\mathbf{q})\,d^{2}\mathbf{p}\,d^{2}\mathbf{q}.
\end{equation}
Making the integration over the momenta and performing a trivial change of variables, we obtain
\begin{equation}
\braket{\psi,\psi'}_{phys}=\int_{\mathbb{R}}\overline{\left[\int_{\mathbb{R}}\phi(q'_{1},q_{2})\,dq_{1}' \right]}\left[\int_{\mathbb{R}}\phi(q''_{1},q_{2})\,dq''_{1} \right]\,dq_{2},  
\end{equation} 
which means that the physical Hilbert space is actually $\mathcal{H}_{phys}=L^{2}(\mathbb{R})$, and therefore the physical states $\psi, \psi'$ belong to $\mathcal{H}_{phys}$. Moreover, $\rho_{phys}(\psi,\psi')$ also satisfies
\begin{equation}
\braket{\psi,\hat{p}_{x}\psi'}_{phys}=\frac{1}{(2\pi\hbar)^{2}}\int_{\mathbb{R}^{4}}\rho_{phys}(\psi,\psi')(\mathbf{p},\mathbf{q})\star p_{x}\,d^{2}\mathbf{p}\,d^{2}\mathbf{q}=0,
\end{equation}
indicating that the constraint is indeed solved by the physical Wigner distribution, as expected.

\subsubsection{Non-relativistic parametrized particle.}

Our second example is devoted to analyze the non-relativistic parametrized  particle. Consider the motion of a particle in three dimensions described by a time independent Hamiltonian $H(\mathbf{q},\mathbf{p})$. This system can be parametrized by adding to the six dimensional phase space a time variable and its corresponding conjugate momentum. Thus, the extended phase space is eight dimensional with local coordinates given by $(q_{0},\mathbf{q},p_{0},\mathbf{p})$. The system is 
characterized by a single first class constraint given by $C(q,p)=p_{0}+H(\mathbf{q},\mathbf{p})=0$, where $q$ and $p$ stand for $(q_{0},\mathbf{q})$ and $(p_{0},\mathbf{p})$, respectively.  As it is well-known, this constraint reduces the fictitious four degrees of freedom to the physical three degrees \cite{Tate}. In order to carry out our quantization program on phase space, 
we will start by considering $\mathcal{H}_{kin}=L^{2}(\mathbb{R}^{4})$ as the kinematic Hilbert space. In accordance with the group averaging approach, the physical Wigner distribution takes the form
\begin{equation}
\rho_{phys}(q,p)=\int_{\mathbb{R}^{4}\times\mathbb{R}}\phi\left(q+\frac{z}{2}\right) \overline{\phi' \left(q-\frac{z}{2} \right)}e^{-\frac{i}{\hbar}zp}\star e_{\star}^{\frac{i}{\hbar}s(p_{0}+H(\mathbf{q},\mathbf{p}))}\,ds\,d^{4}z,
\end{equation}
for $\phi,\phi'\in\mathcal{H}_{kin}$, $z\in\mathbb{R}^{4}$, $s\in\mathbb{R}$ and $ds$ is the invariant measure on $G=\mathbb{R}$. Notice that the 
star-exponential satisfies $e_{\star}^{\frac{i}{\hbar}s(p_{0}+H(\mathbf{q},\mathbf{p}))}=e^{\frac{i}{\hbar}sp_{0}}e_{\star}^{\frac{i}{\hbar}sH(\mathbf{q},\mathbf{p})}$ as  a consequence of the fact that the 
star-product is given by the usual Moyal product (\ref{star}) defined on the phase space $\mathbb{R}^{8}$. In general, the 
star-exponential of any Hamiltonian function can be computed by means of its spectrum as 
\begin{equation}
e_{\star}^{\frac{i}{\hbar}sH(\mathbf{q},\mathbf{p})}=\int_{\mathbb{R}}e^{\frac{i}{\hbar}sE'}\rho_{E^{'}}(\mathbf{q},\mathbf{p})\,dE',
\end{equation}
where $\rho_{E^{'}}(\mathbf{q},\mathbf{p})$ corresponds to the diagonal Wigner distribution associated to an eigenstate of the Hamiltonian with energy $E'$ (for the case of discrete spectra, the integral in the above relation is replaced by a sum), as described in detail 
in~\cite{All}. Making the integration over the time component of the variable $z$, the parameter $s$, and the energy spectrum, we obtain
\begin{equation}
\rho_{phys}(q,p)=\delta(E-p_{0})\rho_{E}(\mathbf{q},\mathbf{p})\star\rho_{p_{0}}(\mathbf{q},\mathbf{p})=\rho_{E}(\mathbf{q},\mathbf{p}),
\end{equation}  
where $\rho_{E}(\mathbf{q},\mathbf{p})$, stands for the standard Wigner distribution (\ref{Wigner}) associated to the energy eigenstates $\phi(\mathbf{q}),\phi'(\mathbf{q})\in L^{2}(\mathbb{R}^{3})$, meaning that the physical Hilbert space is in fact $\mathcal{H}_{phys}=L^{2}(\mathbb{R}^{3})$.  Therefore, 
we notice that the resulting description within the deformation quantization formalism is completely congruent with the quantum theory of the original unconstrained quantum system.

\section{Conclusions}
\label{sec:conclu}
 
In this paper, we have analyzed the quantization of constrained classical systems within the deformation quantization formalism. Following some ideas introduced by the Refined Algebraic Quantization method (RAQ), which corresponds to a refinement of the Dirac quantization program, we obtained the physical Wigner distribution and the physical inner product by means of the star group averaging of the constraints. The star group averaging procedure consists in integrating the star-exponential of the constraints with respect to some appropriate measure and, whenever the integral formulas converge, we explicitly demonstrated that the constraints are solved by the physical Wigner distribution. Therefore, since the physical Wigner distribution corresponds to the phase space representation of the density operator, the physical inner product and the probability densities for observables are obtained by integrating this physical Wigner distribution over the phase space. Finally, we presented a couple of examples in order to illustrate our approach.   
Although at first glance one might think that these examples are too elementary, they are robust enough to show interesting features of our proposal. In the first instance, in the case of a constrained moment, one can immediately recover the integral quantization formulation as developed in~\cite{Gazeau}, while in the second case, for the parametrized particle, the introduced constraint not only reproduces the familiar result~\cite{Henneaux}
but also it 
makes us wonder on the applicability of 
the proposed quantum program to Hamilton-Jacobi type formulations where the constraints are presented in a similar fashion as in this case.
As we have seen, in  both examples the gauge group was identified with the real line.  As future work, we pretend to implement our quantization approach for systems with more general gauge groups of Lie type.  Further, as our initial motivation was the Loop quantum gravity program, 
our intention is to extend our approach to Gauge 
Field theory in order to see the manner in which one may incorporate the physical Wigner distribution in order to describe the quantum properties analyzed within LQG.
This will be done elsewhere.

\section*{Acknowledgments}
The authors would like to acknowledge financial support from CONACYT-Mexico
under projects CB-2014-243433 and CB-2017-283838.

\section*{References}

\bibliographystyle{unsrt}

\end{document}